# *in silico* prediction of protein flexibility with local structure approach.

Tarun J. Narwani[1,2,3,+], Catherine Etchebest[1,2,3,+], Pierrick Craveur[1,2,3,4], Sylvain Léonard[1,2,3], Joseph Rebehmed[1,2,3,5], Narayanaswamy Srinivasan[6], Aurélie Bornot[1,2,3,#], Jean-Christophe Gelly[1,2,3] & Alexandre G. de Brevern[1,2,3,4,*]

[1] INSERM, U 1134, DSIMB, Univ Paris, Univ de la Réunion, Univ des Antilles, F-75739 Paris, France.
[2] Institut National de la Transfusion Sanguine (INTS), F-75739 Paris, France.
[3] Laboratoire d'Excellence GR-Ex, F-75739 Paris, France.
[4] Molecular Graphics Laboratory, Department of Integrative Structural and Computational Biology, The Scripps Research Institute, La Jolla, CA 92037, USA.
[5] Department of Computer Science and Mathematics, Lebanese American University, Byblos 1h401 2010, Lebanon.
[9] MBU, IISc, Bangalore, India
[#] Present address: AstraZeneca, Discovery Sciences, Computational Biology, Alderley Park UK.

*Short title: bioinformatics protein flexibility prediction*

* Corresponding author:
Mailing address: Dr. de Alexandre G. de Brevern, INSERM UMR_S 1134, DSIMB, Université Paris, Institut National de Transfusion Sanguine (INTS), 6, rue Alexandre Cabanel, 75739 Paris cedex 15, France
e-mail : alexandre.debrevern@univ-paris-diderot.fr

## Abstract

Flexibility is an intrinsic essential feature of protein structures, directly linked to their functions. To this day, most of the prediction methods use the crystallographic data (namely B-factors) as the only indicator of protein's inner flexibility and predicts them as rigid or flexible.

PredyFlexy stands differently from other approaches as it relies on the definition of protein flexibility (i) not only taken from crystallographic data, but also (ii) from Root Mean Square Fluctuation (RMSFs) observed in Molecular Dynamics simulations. It also uses a specific representation of protein structures, named Long Structural Prototypes (LSPs). From Position-Specific Scoring Matrix, the 120 LSPs are predicted with a good accuracy and directly used to predict (i) the protein flexibility in three categories (flexible, intermediate and rigid), (ii) the normalized B-factors, (iii) the normalized RMSFs, and (iv) a confidence index. Prediction accuracy among these three classes is equivalent to the best two class prediction methods, while the normalized B-factors and normalized RMSFs have a good correlation with experimental and in silico values. Thus, PredyFlexy is a unique approach, which is of major utility for the scientific community. It support parallelization features and can be run on a local cluster using multiple cores.

The entire project is available under an open-source license at http://www.dsimb.inserm.fr/~debrevern/TOOLS/predyflexy_1.3/index.php.

## 1. Introduction

Proteins are major components of all living cells. They are composed of a succession of amino acids and their structures support directly their functions. Developments of X-ray experiments have increased drastically our knowledge of their inner composition often highlighting the succession of repetitive secondary structures connected by less structured loops. However, this reduced vision lacks to represent the conformational diversity of proteins [1], which is also underestimated by the rather static view provided by X-ray structures. Dynamics information can nevertheless be obtained from X-ray results by considering the Debye-Wallner data, the so-called thermal B-factors, whose accuracy is critically related to the resolution of the structures. However, dynamics may impede the atomic resolution of highly flexible regions, leading to the occurrence of "missing" structural coordinates or be biased by crystal or protein contacts. Hence, alternative approaches are required to bring a complete or at least a more detailed description of protein conformation. For instance, nuclear magnetic resonance (NMR) experiments offer the possibility to get dynamic properties over different timescales but for technical limitations, their use is restricted to proteins with a molecular weight below 40 kDa [2]. Molecular dynamics simulations (MD) are also becoming a powerful tool to get insights in the protein conformational landscape. Unfortunately, simulation timescale still remains, in the best cases, two orders of magnitude below biological timescale. Even fast motions (~hundreds ns) require significant computational time that is not necessarily available in experimental labs. Consequently, to address a simple question like "the location of flexible and rigid regions of a protein" would necessitate to construct an appropriate system and to perform heavy and costly studies. Therefore, the methods that answer this complication are very useful, especially when 3D data are not available, which is a majority case due to the multi-fold increase in metagenomics data. In addition, it would be valuable to obtain quantitative flexibility profile along the sequence. Correlation between normalized B-factors and normalized RMSfs is around 0.45 highlighting the discrepancies between both methods.

Most of the approaches of protein flexibility prediction from the sequence consider only two-state prediction (rigid or flexible) and the flexibility information, as taken from X-ray data. The earlier approaches used simple statistical analyses of B-factor values [3, 4], associated with accessibility predictions [5-7], while more recent approaches use evolutionary information with Artificial Neural Networks (ANNs) [8, 9], Support Vector Regression alone [10] or coupled with Random Forest [11], and of Logistic Regressions [12]. Interestingly, some studies use ANNs with Nuclear Magnetic Resonance data as source of information for protein dynamics with variation of backbone torsion angles [13] or order parameters [14-16] to define the protein flexibility.

In the present paper, we describe the downloadable version of our software, called PredyFlexy that is able to predict flexibility profile from the protein sequence alone through a local 3D structure prediction. PredyFlexy is based on a method previously developed that (i) uses an original description of protein structures with a different view on their flexibility combining both experimental and simulation data (ii) integrates an efficient learning method with evolutionary information. For (i), it considers protein structures as a combination of an ensemble of recurrent structural motifs, called Long Structural Prototypes (LSPs [17, 18]). LSPs are a specific structural alphabet able to approximate every part of the protein structures. LSPs is a library consisting of 120 overlapping structural classes of 11-residues length each, obtained by an unsupervised classification, named Hybrid Protein Model [19]. In this library, relevant sequence-structure relationships were also observed. For (ii) we trained and finely tuned Support Vector Machines (SVM) using observed flexibility information of LSP [20, 21]. Flexibility was considered according to two different descriptors: B-factors from crystallography data and the root mean square fluctuations (RMSF) from molecular dynamics simulations. Using their combination, three classes of flexibility were defined: *rigid*, *intermediate* and *flexible* (see Figure 1 more details). SVMs are powerful supervised learning models with classification algorithms. In a very simple way, an SVM model takes the examples (from test data) as points associated to a feature (label) and mapped these data to a higher dimension to classify the examples into different labels. It must be noticed that PredyFlexy was trained on the data coming from ordered protein structures (providing normalized B-factors), each one being simulated with MDs (providing normalized RMSF); thus the prediction only used the amino acid sequence to predict these flexibility features.

In details, the method works in two steps: first, it predicts the local structures (LSPs) along the sequence and then it predicts the flexibility of the sequence, based on the observed flexibility of the predicted LSPs. The prediction rate is slightly better than the one of PROFbval [8, 9] optimized for two classes. Importantly, PredyFlexy also propose a confidence index for assessing the quality of the prediction rate. PredyFlexy software is available under an open-source license at http://www.dsimb.inserm.fr/~debrevern/TOOLS/predyflexy_1.3/index.php.

## 2. Materials and Methods

As mentioned above, the main idea of PredyFlexy is to use a large set of protein dynamics to have access to their flexibility through normalized Root Mean Square Fluctuations (RMSFs) obtained from MD simulations and experimental B-factors (normalized). The principle is to define three classes of flexibility using these two flexibility descriptors, in order to reduce the bias introduced by either one, e.g. crystal contact for X-ray structures [22] and inadequate sampling of conformational states for MD simulations [23]. Then 120 SVMs are used to predict LSPs. From this prediction, the 5 most probable LSPs, on each position along the sequence, are used to predict the flexibility as *rigid*, *intermediate,* and *flexible*. Prediction rate is slightly better that the well-established PROFbval that only predicts two classes, which is simpler but less accurate description.

Importantly, two other values are predicted, the normalized B-factors and the normalized RMSFs profile along the sequence. Finally, a confidence index (CI) is proposed. The CI is based on the discriminative power of the SVM classifiers. CI is graded from 1, easy-to-predict regions (high confidence index), to 19, very difficult to predict (low confidence index).

***2.1 Data sets and molecular dynamics***. A primary dataset of 40 proteins was used previously [20, 21]. A novel databank of 169 X-ray structures (~4 times than previously used), taken from Protein DataBank (PDB) was tested here. It was extracted using ASTRAL 2.03 at 40% sequence identity. The dataset was filtered out based on structure resolution better than 1.5 Å, and without presence of heteroatoms (other than water), alternate or modified residues in the chain. Only globular proteins, with chain length ranging between 50 and 250 residues, were selected. In-house parser was used to filter out and to fetch the information. The 169 domains represent a rather equilibrated repartition among the different SCOP classes: all-α represents 18.9% of the chains, all-β represents 29.6%, α/β: 24.8%, and α+β: 26.7%. Three molecular dynamic (MD) simulations were performed for each protein structure with GROMACS 4.5.7 software using AMBER99sb force field.

***2.2. Implementation.*** PredyFlexy 1.3 is composed of 9 successive steps presented in Figure 2. It relies on the use of in-house Python scripts and three external tools.

These tools are (in order of usage): BLAST version 2.2.09 with the dedicated databank (Swissprot 2006), svm_classify from SVM-light Version 6.01 and svm-scale from libsvm version 2.81 (see Discussion section). All tools can be used with classical distribution such as Ubuntu 16.04 or 18.04, recent RedHat, as for MacOS (tested with 10.13.6). For MacOS, *virtualenv* can be useful in case of another PSI-BLAST installation and different databases.

The steps:

(1) A sequence is read with only amino acids (fasta format).

(2) PSI-BLAST is used over four iterations.

(3) It generates PSSM.

(4) This PSSM is normalized to be used by SVMs.

(5) PSSM is split into sub matrices of window size 11-21.

(6) Each of the sub matrices is used as an input for each of the 120 SVMs. As it is the most time consuming step, multi threading can be applied.

(7) The scores from each SVM are ranked at each position allowing the prediction of LSPs.

(8) The flexibility prediction in three classes is performed, simultaneously with the prediction of normalized B-factors and normalized RMSFs. Confidence index is evaluated at each position

(9) Outputs are written as shown in Figure 3.

***2.3. Installation***. After downloading the archive (predyflexy1.3.tar.gz) and the proper tools (they can be found on the same website if needed), uncompress the archive (using the following command "tar -xzvf predyflexy1.3.tar.gz"). Then, the three external tools (e.g., BLAST, svm_classify and svm-scale) must be installed; more instructions are provided in the README. The external tools relied on Linux and therefore only Linux version of the software is supported.

To insure that the installation was done correctly, a test mode exists. It consists in running a specific sequence for which the results have been pre-computed. Each intermediate and final result is compared with it.

The test is done using "python pred.py -t", while the sentence "Congratulations PredyFlexy works correctly" shows that the installation was successfully done.

***2.4. Usage***. A large number of options exists (see Additional file 1). They can be considered in three major cases. The first one concerns the directory for the installed tools, some default paths are provided but you can modify it according to the location you choose. The second one concerns parameters for the external tools; they must not be modified (they are here for development purposes only). The last series of options concern the output results, user can modify it without any trouble, e.g. to do only a prediction of LSPs and not flexibility.

A simple example is: "python pred.py -f DATA/TEST/Prot_0.fasta --confidence --flex", the whole prediction will be done. Some interesting options are -cpu to use a maximum number of CPUs, and -DIR which will be the new output directory, otherwise the default is a random directory which is created, e.g. TMP18196054366918182720.

***2.5. Understanding the output***. Figure 3 shows the classical output of PredyFlexy. The 10 first and 10 last residues are not predicted due to the size of the sequence window, i.e. the sequence window is of size 21 centred on the $11^{th}$ residue.

The output is in a very simple format. The columns are in order: the position of the residue, its type, the five best LSPs, the prediction information with the confidence index, the flexibility prediction, the predicted normalized RMSFs and the predicted normalized B-factors.

## 3. Results & Discussion

In a recent review, we have underlined the interests of local protein structure conformations, namely the structural alphabets, in the analysis of protein flexibility [1]. In this field, PredyFlexy WebServer [21] remains a powerful approach. Prediction was assessed using classical measures such as $Q_3$ and F-measure that combines accuracy and coverage, showing the correct prediction rates without particular bias. Since the original methodological publication [20], many users have requested for a standalone PredyFlexy tool because PredyFlexy webserver does not allow performing predictions on a large dataset of protein sequence. Considering this demand, we decide to make available a downloadable package, with (new) options that the user can choose to modify. Additionally, since the publication of the original paper, we have significantly enlarged our MD dataset, *i.e.* 169 new SCOP domains instead of 54. We have also increased the simulation time (150 ns *per* protein, with three independent replicates, *i.e.* 3 times more proteins with longer times than before). The MD protocol was the same as in the original paper (see [20, 24] for details). These new data are used to evaluate the stability of the prediction quality. Pearson correlation between normalized RMSFs and normalized B-factors equals 0.43, which is comparable to the previous value calculated with a more limited dataset [20]. Table 1 provides the results of the prediction of the protein for the new dataset; it remains highly consistent with previous results. The number of problematic cases, *i.e.* error between flexible and rigid region diminish by 2%, while the prediction of rigid position increases from 47.4% to 51.5% and of flexible positions from 55.0% to 61.5%. An important point is the very limited number of cases for which flexible fragments are predicted as rigid and vice-versa (false positives), they represent less than 8% of the cases. It also shows the high efficiency of PredyFlexy v1.3 to give an accurate, quantitative evaluation of the flexibility without the necessity to perform long simulations. Indeed, for the set of proteins tested, the results were obtained in a few hours, while the MD simulations required many weeks (months) of calculations on a high performance cluster. This result considerably strengthens the interest of PredyFlexy compared to other heuristic methods.

This study also allowed us to identify a critical issue related to PSI-BLAST tools. Indeed, when we used a newer or the latest version for PSI-BLAST, the results were different. Actually, only 45% of the hits found with the default version 2.2.09 were obtained with the latest BLAST versions. We tried to optimize different program options but never succeeded to have recruited more than 5% sequences. Therefore, changes of the versions were critical leading to very different outputs, and as the Support Machines with it. Thus the results were quite differently both for LSPs prediction and flexibility prediction. It is so well advised to keep the same configuration.

Finally, this work gives us the opportunity to revisit flexibility literature and to assess the performance of PredyFlexy compared to the most recent tools. Table 2 summarizes the different protein flexibility prediction methodologies. It is striking to see that the number of such tools is limited. The importance taken by disorder prediction methods since many years might have contributed to this lack of interest in flexibility prediction tools. For instance, the recent DynaMine approach (webserver: http://dynamine.ibsquare.be/, download: http://dynamine.ibsquare.be/download/) uses backbone flexibility at the residue-level in the form of backbone N-H $S^2$ order parameter values, i.e. these parameters comes from NMR experiments and gives direct information on the motion of these bonds. Nonetheless, it focuses on the prediction of disordered regions with accuracy comparable to the most sophisticated existing disorder predictors [25, 26].

Only two other approaches are available as local tools: PROFbval with a Debian package [8, 9] and RCI with one python file [14, 15]. PROFbval is based on B-factors and propose only a two class prediction (flexible or rigid). Some proteins were tested online and provides results in correspondence with the first dataset [20]. With only two states to predict, PROFbval reach 53.3% and 71.9% depending on the threshold between rigid and flexible region. Similar results are reached with our approach when it is adapted to the same two classes (it must be noticed that PredyFlexy has been defined for 3-states and not 2). RCI is based on NMR data and needs NMR, i.e. cannot take a simple amino acid sequence. Moreover, the learning was done on a very small set of proteins (14 proteins) and, only a flexibility profile is provided. Hence, PredyFlexy software is the most recent approach and the only one that proposes three different flexibility predicted states, a predicted B-factor profile, a predicted RMSF profile, and a confidence index.

The Figure 4 shows the execution time of the PredyFlexy for three proteins with increasing residue length (150, 250 and 550). Computation time is mainly function of protein length, and over 90% is spent in the SVMs steps. Over a perfect theoretical distribution (see dotted line), there is a loss of a third with four CPUs, as there are over hundred SVMs, the communication time is a limiting factor that may not be easily reduced. In view of the benchmark presented here (and others), we propose to use four CPUs (if possible) for best computational efficiency.

## 4. Conclusion

The state-of-the-art on protein flexibility prediction was based on B-factors, directly from X-ray structures experiments. However, this flexibility measure is mainly linked to uncertainty of the atoms positions and can be biased by the experimental conditions. The addition of a second opinion, i.e. MDs, is pertinent to provide a better view of inner dynamics through RMSf. We have therefore, linked both of the results taken into account from X-ray data and MDs. For the first time, not two classes of flexibility, namely flexible and rigid, were defined but we added an intermediate one, leading to 3-classes. The results are interesting as with the 3 classes prediction provides slightly better resolution into the flexibility than the 2-classes classification, and the case of problematic confusion is highly limited, i.e. predicting a rigid part when it is flexible or inversely were seen. Moreover, PredyFlexy software is the only methodology offering the prediction of B-factor profile and RMSF profile; also we strongly believed in the interest of confidence index to help the expert and non-expert users.

This research leads to new questions. Some interesting ones concern the most widely used Structural Alphabet, the Protein Blocks [27, 28] and their behaviours. A general question remains with the precise delineation (that does not exist) between flexibility and disorder, and for instance with the existence of specific cases as Dual Personality Fragments [29], found both in ordered and disordered states. It is also strongly linked to the bridge between dynamics and flexibility as we have recently shown with specific case of helices shift between $\alpha$-helix, $\pi$-helix, and $3_{10}$-helix. Taken from X-ray structures, we showed that a limited number of them were in reality not stable and associated with very flexible regions [30].

## Acknowledgments

We would like to thank Dr. Nicolas Shinada, Dr. Tatiana Galochkina, and Dr. Akhila Melarkode Vattekatte for fruitful discussions. This work was supported by grants from the French Ministry of Research, University of Paris Diderot – Paris 7, French National Institute for Blood Transfusion (INTS), French Institute for Health and Medical Research (INSERM). AdB, NS, and TJN also acknowledge the Indo-French Centre for the Promotion of Advanced Research / CEFIPRA for collaborative grants (numbers 5302-2). This study was supported by grants from the Laboratory of Excellence GR-Ex, reference ANR-11-LABX-0051. The labex GR-Ex is funded by the programme “Investissements d’avenir” of the French National Research Agency, reference ANR-11-IDEX-0005-02. Calculations were performed on an SGI cluster granted by Conseil Régional Ile de France and INTS (SESAME Grant). Research in NS group is supported by Mathematical Biology program and FIST program sponsored by the Department of Science and Technology and also by the Department of Biotechnology, Government of India in the form of IISc-DBT partnership programme. Support from UGC, India – Centre for Advanced Studies and Ministry of Human Resource Development, India is gratefully acknowledged. NS is a J. C. Bose National Fellow. The authors were granted access to high performance computing (HPC) resources at the French National Computing Center CINES under grants no. c2013037147, c2016077621 and A0010707621 funded by the GENCI (Grand Equipement National de Calcul Intensif).

The funding bodies have no role in the design of the study and collection, analysis, and interpretation of data and in writing the manuscript

*Authors' contributions*. CE and AdB jointly conceived the entire protein flexibility prediction study. AB had done the first experiments for protein structure analyses, molecular simulations and prediction optimization. JR and AdB jointly conceived the second series of molecular dynamics simulations that was performed by PC and JR. AdB had rewritten the code in Python, with the help of SL and JCG. TN had optimized it as the installation process. TN had performed all the tests for the second series of molecular dynamics simulations. CE, NS, JCG and AdB had supervised the study. CE and AdB wrote the main paper. All authors discussed the results and implications and commented on the manuscript at all stages.

## Figure legends

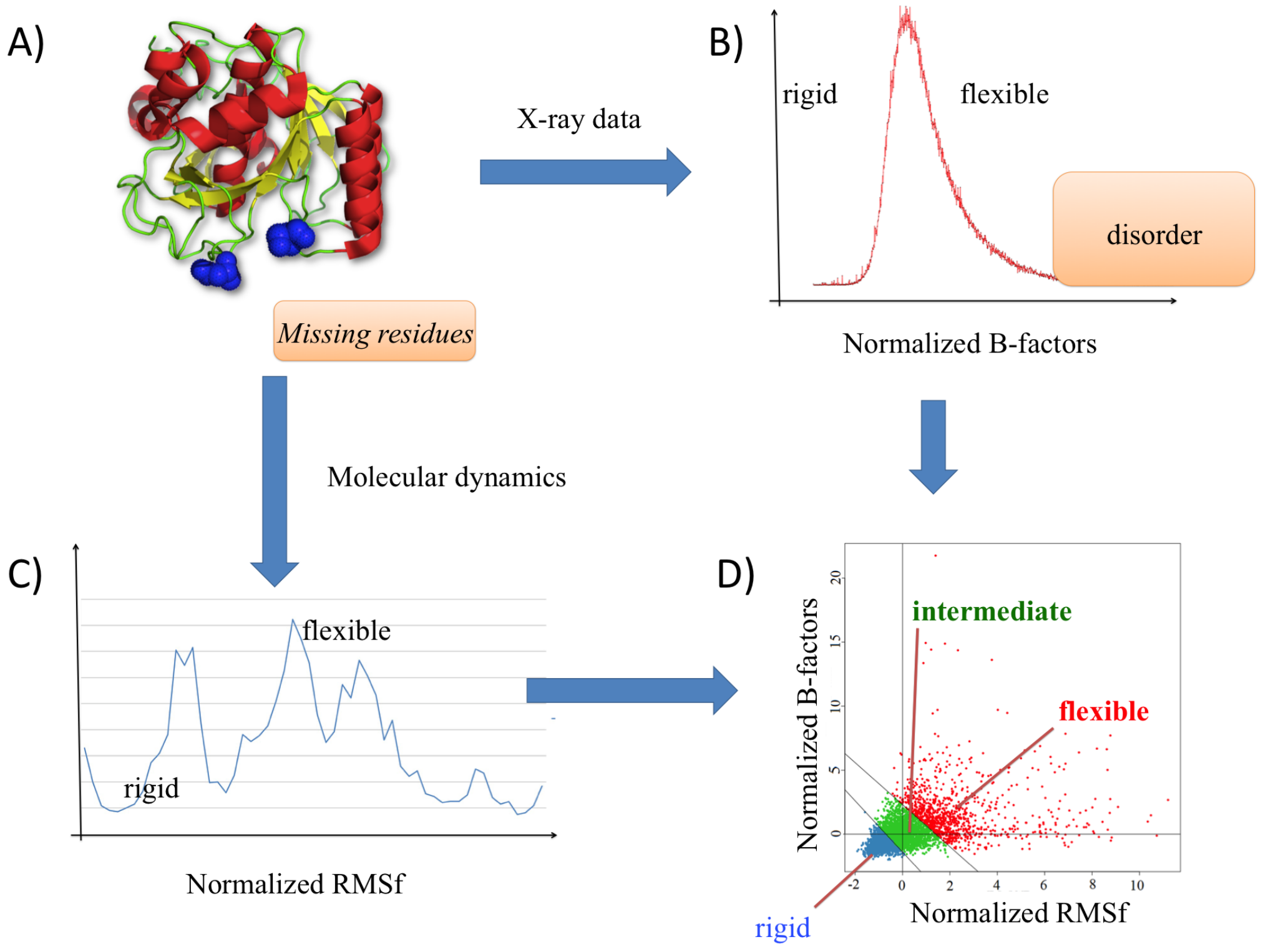

**Figure 1.** *Generation of the structural data*. (A) A dataset of protein structures was selected; they were of high quality with no missing residues. (B) From the X-ray structures were extracted the B-factors, they defined classically rigid regions (low normalized B-factors) and flexible regions (high normalized B-factors), they characterized ordered regions, while disorder (not taken into account here) is characterized by the missing residues. (C) For each protein structures, molecular dynamics simulations were performed and RMSf were computed; they defined classically rigid regions (low normalized RMSf) and flexible regions (high normalized RMSf). (D) The correlation between normalized B-factors and normalized RMSf is often close to 0.45 as shown here. From this distribution, we defined two series of threshold characterizing rigid (blue), intermediate (green) and flexible regions (red).

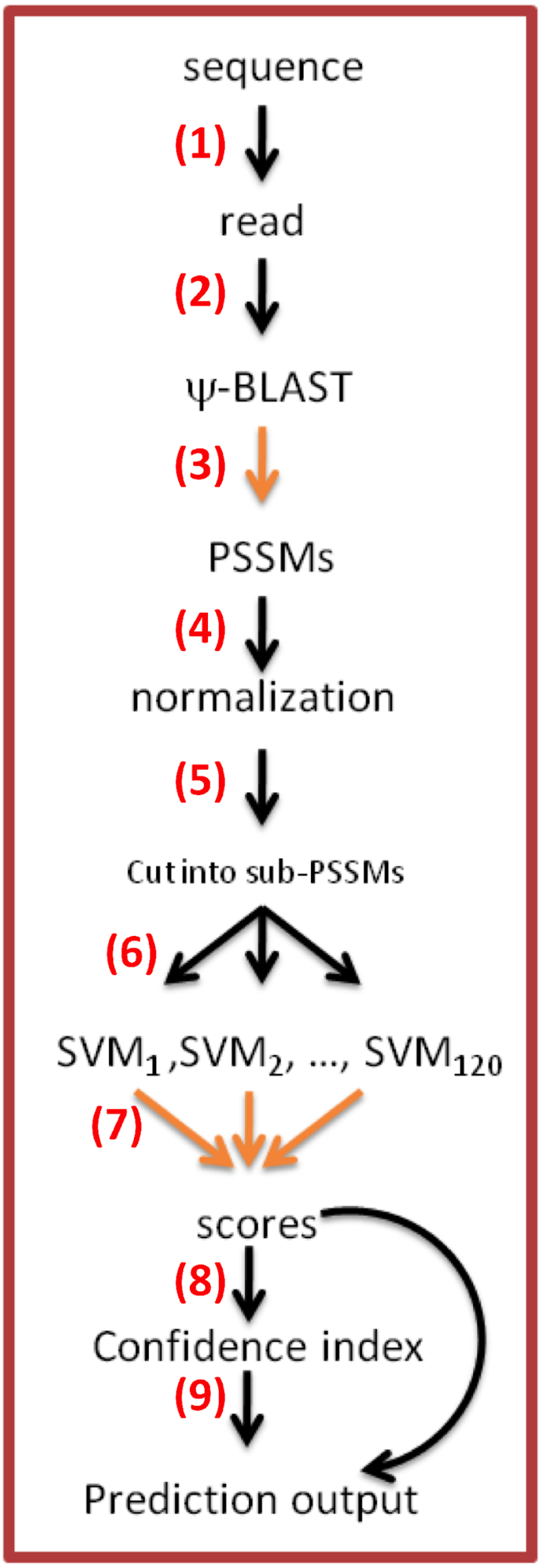

**Figure 2.** *The framework of PredyFlexy and underlying methods*. (1) User must give a single sequence as (2) input. (3) A PSSM is computed using PSI-BLAST, (4) a normalization step is done and then (5) the normalized PSSM is split into fragments of 11 residues. (6) A specific prediction of LSPs is done using trained SVMs scores. As this step is time consuming, parallelization is interesting to perform, and then the 120 predicted LSPs are ranked in regards to their LSP values. (7) The best five are kept, prediction of flexibility relies on them (8-9) providing (a) in three states (rigid, intermediate and flexible), and (b) predicted B-factor$_{Norm}$, and (c) RMSF$_{Norm}$. A confidence index is also given.

```
                 identifiers
                 |
>1OMHA_0  <---/
  [...]
                    <--- first aa are not predicted

     ---amino acids          -- confidence index
    /                        /
  11 G  47  93   4   3  75  15   1  0.979  0.522
  12 L   4  34  79 114  58  13   0  0.130 -0.027
  13 R  92  96  95  57  89  15   0  0.209  0.080
  14 F  57  96  95 106  55  13   0 -0.183 -0.367
  15 I  79  56  58 107   8  15   0 -0.390 -0.415
  16 D  57 107  55  43  95  13   0  0.032 -0.131
  17 L  11  79  12  64  44  10   0 -0.004  0.025
  18 F  42  11  12 110 117  10   0 -0.011  0.195
   |     |   |   |   |    |      |      \      /
  /      \   \   |   /    /     /        \   /
number    \   \  |  /    /   flexibility  flexibility
           -5 prediced --     index         profils
             prototypes                   (RMSf and B-factors)
```

**Figure 3.** *Protein prediction example.* The first line corresponds to the identifier provided by the user. Then the results is provided per columns with (i) the residue number, the first amino acids being noted +1, (ii) the corresponding amino acids, (iii) the 5 best predicted LSPs, they will dictated the flexibility prediction, (iv) the confidence index ranging between 1 (excellent prediction) to 19 (very poor prediction), (v) the predicted flexibility index based on experiments and simulations, with (0) rigid, (1) intermediate and (2) flexible, (vi) the two flexibility profiles with predicted normalized RMSf and predicted normalized B-factors.

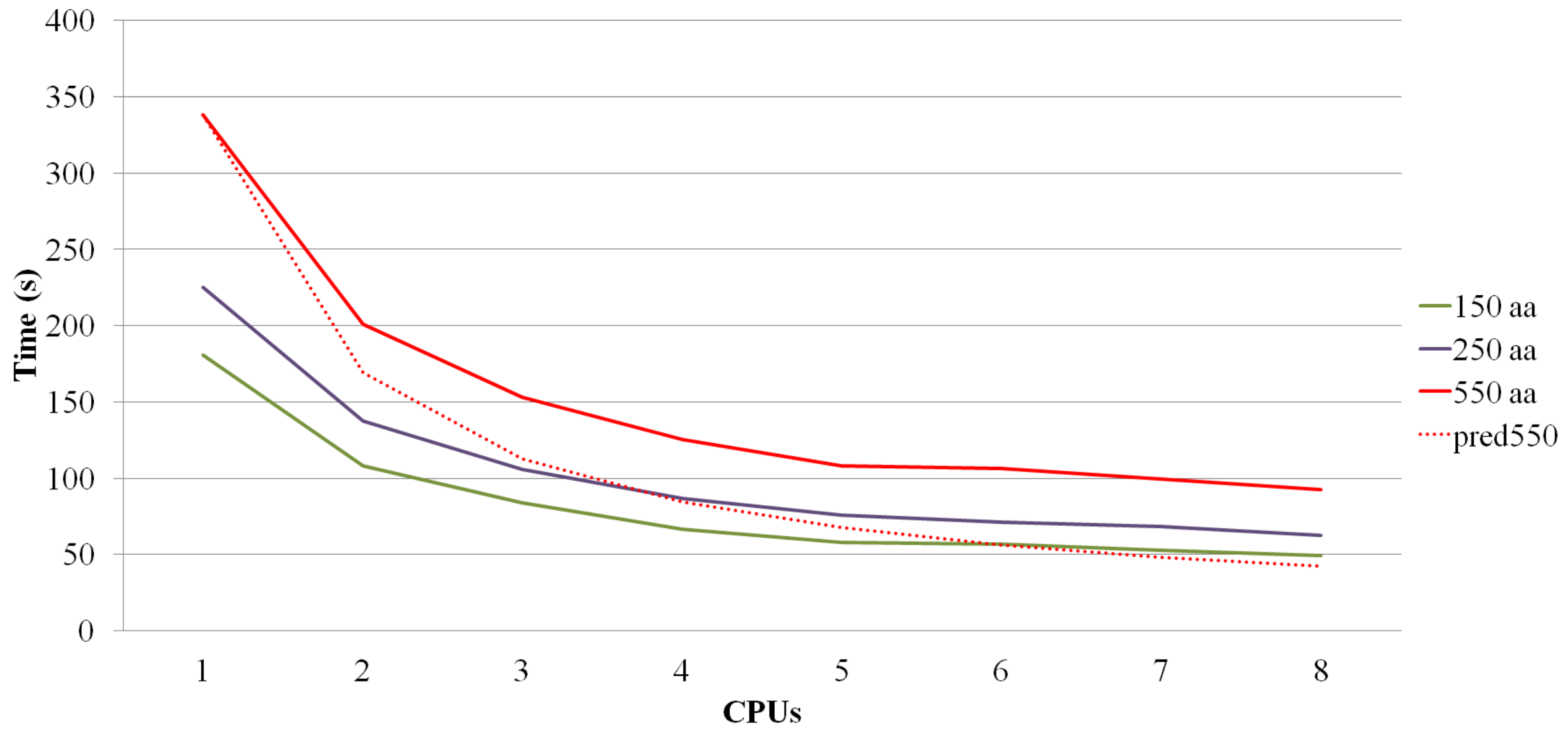

**Figure 4**. Evolution of computation times in regards to the number of CPUs, for three examples (150, 250 and 550 residues, in green, violet and red colours), shown in dotted red an optimal computation time for the last example. The computer use is a classical PC with 8 nodes (x86 64 bits with 8 Mo of RAM, Ubuntu 14.04).

## Additional File

**Additional file 1 [pdf]**. **Additional file 1**. *The different options of PredyFlexy software*.

| Obs vs Pred (%) | Rigid | Intermediate | Flexible | Sum |
|---|---|---|---|---|
| Rigid | **51.55** | 37.33 | 11.11 | 100.00 |
| Intermediate | 19.85 | **43.97** | 36.18 | 100.00 |
| Flexible | 4.90 | 33.61 | **61.49** | 100.00 |

**Table 1.** *PredyFlexy prediction on the new 169 proteins*. Is provided the distribution of prediction for the three classes of flexibility. Only 11% of rigid fragment are predicted as flexible and only 5% of flexible fragments are predicted as rigid.

| Name | Ref | url | year | download |
|---|---|---|---|---|
| *Schulz* | [3] | *NA* | 1985 | *NA* |
| *Riiikonen* | [4] | *NA* | 1994 | *NA* |
| FLePS | [5] | http://bioinf.modares.ac.ir/software/fleps/index.html | 2001 | *NA* |
| *Teasdale* | [10] | *NA* | 2005 | *NA* |
| PROFbval | [8-9] | https://www.predictprotein.org | 2005 / 2006 | https://rostlab.org/owiki/index.php/PROFbval<br>Debian package |
| RCI | [14-15] | http://randomcoilindex.com/cgi-bin/rci_cgi_current.py | 2005 / 2007 | http://wishart.biology.ualberta.ca/download/rci/<br>one python file |
| FlexRP | [12] | *NA* | 2007 | *NA* |
| PredictS$^2$ | [16] | *NA* | 2008 | *http://www.palmer.hs.columbia.edu/software/predictS2*<br>*compiled software* |
| FlexPred | [6-7] | http://flexpred.rit.albany.edu | 2008 / 2008 | *NA* |
| PredBF | [11] | http://www.csbio.sjtu.edu.cn/bioinf/PredBF/# | 2009 | *NA on 27-07-2016* |
| *Zhou* | [13] | *NA* | 2010 | *NA* |
| PredyFlexy | [19-20] | http://www.dsimb.inserm.fr/dsimb_tools/predyflexy | 2011 / 2012 | http://www.dsimb.inserm.fr/~debrevern/TOOLS/predyflexy_1.3/index.php<br>python and external tools |

**Table 2.** *Flexibility prediction method summary*. Are provided by chronological order the different methodologies with references, url, year of publications and the url for download of the methods.

```
Please note an important number of options (some are dedicated to blast, ...):


Usage: pred.py -f the-fasta-file [options]

Options:
  -h, --help            show this help message and exit
  -f FILENAME, --file=FILENAME
                        obligatory                               [no default]
  -o OUTNAME, --output=OUTNAME
                        the name of the files              [default = TMPxxx]
  -D DIROUTPUT, --DIR=DIROUTPUT
                        the temporary directory              [default = TMP/]
  -c CPU_NUMBERS, --cpu=CPU_NUMBERS
                        number of cpus to be used             [default = all]
  --confidence          to compute the confidence index          [default=NO]
  --flex                prediction of flexibility                [default=NO]
  -t, --test            test                                [default=NO] [na]
  --PSIBLASTDIR=PSIBLAST_DIRECTORY
                        directory of psi-blast       [default: SOFTS/psiblast]
  --PSIBLASTdb=PSIBLAST_DB
                        databank of psi-blast (formatted)
                        [SOFTS/db]
  --PsBround=NB_ROUND   psi-blast number of round               [default = 4]
  --PsBeval=E_VALUE     psi-blast e-value                  [default = 0.0001]
  --PsBval=PSB_VALUE    psi-blast value                    [default = 0.0001]
  --Makemat=MAKEMAT_DIRECTORY
                        directory of makemat          [default: SOFTS/makemat]
  --GCM=GETMAT_DIRECTORY
                        directory of get_clean_matrix        [default: SOFTS/]
  --SVMclassify=SVMCLASSIFY_DIRECTORY
                        directory of SVM classify            [default: SOFTS/]
  --SVMscale=SVMSCALE_DIRECTORY
                        directory of SVM scale               [default: SOFTS/]
  -q, --quiet           say nothing                             [not default]
  -v, --verbose         verbose                                     [default]
  -w, --very-verbose    terribly verbose                        [not default]
  -n, --noban           No ban                                  [default: on]
  --firefox             firefox                                 [not default]

Default parameters have been optimized and corresponds to the results of the papers [1, 2, 3].
```

**Supplementary material 1**. *The different options of PredyFlexy software*. (see text for explanations).